\documentclass[letterpaper, 10 pt, conference]{ieeeconf}  % Comment this line out if you need a4paper

\IEEEoverridecommandlockouts                              % This command is only needed if
\usepackage{amsmath}
\usepackage{amssymb} % for real number R:
\usepackage{bm}
\usepackage{bbold}
\usepackage{cite}
\usepackage{url}
\usepackage{balance}

\usepackage{tikz}
\usetikzlibrary{arrows,matrix,positioning,patterns}
\usetikzlibrary{calc}
\usetikzlibrary{plotmarks}
\usepackage{pgfplots} % to include tikz-figs.
\usetikzlibrary{external}

\newtheorem{rmk}{Remark}

\usepackage{amsmath}
\usepackage{amssymb} % for real number R:

\usepackage{algorithm} %ctan.org\pkg\algorithms
\usepackage{algpseudocode}

\title{\LARGE \bf Trajectory Planning Under Vehicle Dimension Constraints\\ Using Sequential Linear Programming}

\author{Mogens Graf Plessen$^{*,1}$, Pedro F. Lima$^{*,2}$, Jonas M{\aa}rtensson$^{2}$, Alberto Bemporad$^{1}$, and Bo Wahlberg$^{2}$% <-this % stops a space
\thanks{*The authors contributed equally when MGP was visiting KTH.}% <-this % stops a space
\thanks{$^{1}$MGP and AB are with IMT School for Advanced Studies Lucca, Piazza S. Francesco 19, 55100 Lucca, Italy, {\tt\small{mogens.plessen, alberto.bemporad}@imtlucca.it}.}
\thanks{$^{2}$PFL, JM, and BW are with Integrated Transport Research Lab and ACCESS Linnaeus Centre, Department of Automatic Control, KTH Royal Institute of Technology, SE-10044 Stockholm, Sweden, {\tt\small {pfrdal, jonas1, bo}@kth.se}.}%
}

\renewcommand{\baselinestretch}{0.895}
\begin{document}

\maketitle
\thispagestyle{empty}
\pagestyle{empty}

%%%%%%%%%%%%%%%%%%%%%%%%%%%%%%%%%%%%%%%%%%%%%%%%%%%%%%%%%%%%%%%%%%%%%%%%%%%%%%%%
\begin{abstract}

This paper presents a spatial-based trajectory planning method for automated vehicles under actuator, obstacle avoidance, and vehicle dimension constraints. Starting from a nonlinear kinematic bicycle model, vehicle dynamics are transformed to a road-aligned coordinate frame with path along the road centerline replacing time as the dependent variable. Space-varying vehicle dimension constraints are linearized around a reference path to pose convex optimization problems. Such constraints do not require to inflate obstacles by safety-margins and therefore maximize performance in very constrained environments. A sequential linear programming (SLP) algorithm is motivated. A linear program (LP) is solved at each SLP-iteration. The relation between LP formulation and maximum admissible traveling speeds within vehicle tire friction limits is discussed. The proposed method is evaluated in a roomy and in a tight maneuvering driving scenario, whereby a comparison to a semi-analytical clothoid-based path planner is given. Effectiveness is demonstrated particularly for very constrained environments, requiring to account for constraints and planning over the entire obstacle constellation space.

\end{abstract}

%%%%%%%%%%%%%%%%%%%%%%%%%%%%%%%%%%%%%%%%%%%%%%%%%%%%%%%%%%%%%%%%%%%%%%%%%%%%%%%%
\section{Introduction}

Automated vehicles can address various challenges. Fuel consumption can be reduced by means of platooning~\cite{al2010experimental}, and anticipative driving in car-2-car and car-2-infrastructure communicating environments~\cite{homchaudhuri2017fast,gozalvez2012ieee}. Traffic safety may be increased by means of automated handling of vehicles at their friction limits~\cite{massera2016guaranteed,altche2017high,funke:lanechangeEP}. Congestion in cities can be reduced by means of coordinated traffic flows~\cite{hult2016coordination}. We can distinguish between longitudinal and steering-related vehicle control. The former is much simpler when considered isolatedly and it is introduced commercially~\cite{BMWtrafficjam}. Steering-applications are more complicated, since the exact traveling trajectory is decisive for permissible traveling speeds within friction limits, thereby affecting vehicle stability. In general, we can distinguish between high- and low-velocity driving scenarios. For the former, steering is relevant for obstacle avoidance and throughput maximization on highways with vehicles of different agility capabilities~\cite{plessen2016multi}. For the latter, steering is relevant for tight maneuvering.
\begin{figure}[t]
\vspace{0.0cm}\centering
\begin{tikzpicture}[thick,scale=0.6, every node/.style={scale=0.6}]%§
% define variables.
\coordinate (R) at (1.8,1.55);
\coordinate (F) at ($ (R) + 4*({cos(30)},{sin(30)})$);
\coordinate (C) at ($ (R) + 1.667*({cos(30)},{sin(30)})$);
\coordinate (Ra) at ($ (C) - 2.6*({cos(30)},{sin(30)})$);
\coordinate (Raa) at ($ (C) - 2.92*({cos(30)},{sin(30)})$);
\coordinate (Rb) at ($ (C) + 3.6*({cos(30)},{sin(30)})$);
\def\psis{18} % for road (psi_s).
\def \wheelL {0.5}
\def \lineRhoSTop {2.2}
\def \lineRhoSBottom {1.7}
% coord. sys.
\draw[->] (0, 0) -- (C|-,0);
\draw[->] (0, 0) -- (0, |-C);
\node[color=black] (a) at ($ (C|-,0) + 0.25*({cos(0)},{sin(00)})$) {$x$}; % xlabel
\node[color=black] (a) at ($ (0, |-C) + 0.25*({cos(90)},{sin(90)})$) {$y$}; % ylabel
% straight line
\draw[color=black] (R) -- (F);
% frame
\draw[rounded corners,black,fill=green!40,fill opacity=0.2,rotate around={30:(C)}] ($ (C) - ({2.6},{0.75})$) rectangle ($ (C) + ({3.6},{0.75})$);
% parallel straight line
\draw[color=black,|-|] ($ (Ra) - 1.1*({cos(-60)},{sin(-60)})$) -- ($ (C) - 1.1*({cos(-60)},{sin(-60)})$) node [midway, above] {$a$};
\draw[color=black,-|] ($ (C) - 1.1*({cos(-60)},{sin(-60)})$) -- ($ (Rb) - 1.1*({cos(-60)},{sin(-60)})$) node [midway, above] {$b$};
\draw[color=black,|-|] ($ (Raa) + 0.75*({cos(-60)},{sin(-60)})$) -- ($ (Raa) - 0.75*({cos(-60)},{sin(-60)})$) node [midway, left] {$2w$};
% rear wheel
\draw[rounded corners,black,fill=black,fill opacity=0.2,rotate around={30:(R)}] ($ (R) - \wheelL*({cos(0)},{0.2*sin(90)})$) rectangle ($ (R) + \wheelL*({cos(0)},{0.2*sin(90)})$);
% front wheel
\draw[rounded corners,black,fill=black,fill opacity=0.2,rotate around={60:(F)}] ($ (F) - \wheelL*({cos(0)},{0.2*sin(90)})$) rectangle ($ (F) + \wheelL*({cos(0)},{0.2*sin(90)})$);
% center of gravity.
\fill (C) circle [radius=2pt];
% continuation cog.
\draw[color=black!50] (C) -- ($ (C) + 1.55*({cos(0)},{sin(0)})$);
% v_x.
\draw[->,color=red!50,>=latex',ultra thick] (C) -- ($ (C) + 1*({cos(30)},{sin(30)})$);
\node[color=black] (a) at ($ (C) + ({0.4*cos(30)},{0.47*sin(90)})$) {$v$};
% rho-radius (road)
\fill[color=blue!70] ($ (C) + \lineRhoSTop*({cos(90+\psis)},{sin(90+\psis)})$) circle [radius=1.5pt];
\draw[color=blue,->] ($ (C) + \lineRhoSTop*({cos(90+\psis)},{sin(90+\psis)})$) -- ($ (C) + \lineRhoSBottom*({cos(-(90-\psis))},{sin(-(90-\psis))})$) node [pos=0.11, left=0.0] {$\rho_s(s)$} node [pos=0.97, left=0.] {$s$};
% arc rho-radius
%\draw[color=black!50,->] ($ (C) + 4*({cos(90+\psis)},{sin(90+\psis)}) + 7.5*({cos(-93)},{sin(-93)})$) arc (-93:-40:7.5); % orig.
\draw[color=black!50,->] ($ (C) + \lineRhoSTop*({cos(90+\psis)},{sin(90+\psis)}) + {\lineRhoSTop+\lineRhoSBottom}*({cos(-93)},{sin(-93)})$) arc (-93:-40:{\lineRhoSTop+\lineRhoSBottom});
\node[color=black!50] (a) at ($ (C) + (3.6,-0.25)$) {road centerline};
%% arc \dot{\psi}_s
%\draw[color=blue,->] ($ (C) + \lineRhoSTop*({cos(90+\psis)},{sin(90+\psis)}) + 0.3*({cos(180)},{sin(180)})$) arc (180:330:0.3);
%\node[color=blue] (a) at ($ (C) + \lineRhoSTop*({cos(90+\psis)},{sin(90+\psis)}) + 0.6*({cos(330)},{sin(330)})$) {$\dot{\psi}_s$};
% dashed tangent to road
\draw[-,color=black!50, dashed] ($ (C) + \lineRhoSBottom*({cos(-(90-\psis))},{sin(-(90-\psis))}) - 2*({cos(\psis)},{sin(\psis)}) $) -- ($ (C) + \lineRhoSBottom*({cos(-(90-\psis))},{sin(-(90-\psis))}) + \lineRhoSTop*({cos(\psis)},{sin(\psis)}) $) node [pos=0.95, right=0.1] {tangent};
%% \dot{s}
%\draw[->,color=blue,>=latex',ultra thick] ($ (C) + \lineRhoSBottom*({cos(-(90-\psis))},{sin(-(90-\psis))})$) -- ($ (C) + \lineRhoSBottom*({cos(-(90-\psis))},{sin(-(90-\psis))}) + 1.5*({cos(\psis)},{sin(\psis)}) $) node [pos=0.95, left=0.5] {$\dot{s}$};
% dashed tangent to road shifted parallel through CoG
\draw[-,color=black!50, dashed] ($ (C) - 3*({cos(\psis)},{sin(\psis)}) $) -- ($ (C) + 3*({cos(\psis)},{sin(\psis)}) $);
% arc delta
\draw[-,color=black!50, dashed] ($ (F)$) -- ($ (F) + 1*({cos(30)},{sin(30)}) $);
\draw[-,color=black!50, dashed] ($ (F)$) -- ($ (F) + 1*({cos(60)},{sin(60)}) $);
\draw[->, color=blue] ($ (F) + 0.61*({cos(30)},{sin(30)})$) arc (30:60:0.65);
\node[color=blue] (a) at ($ (F) + 0.77*({cos(45)},{sin(45)})$) {$\delta$};
% arc e_psi
\draw[->, color=blue] ($ (C) + 1.25*({cos(\psis)},{sin(\psis)})$) arc (\psis:30:1.25);
\node[color=blue] (a) at ($ (C) + 1.5*({cos(22)},{sin(22)})$) {$e_\psi$};
% arc psi_s close to (C)
\draw[->, color=blue] ($ (C) + 1.25*({cos(0)},{sin(0)})$) arc (0:\psis:1.25);
\node[color=blue] (a) at ($ (C) + 1.5*({cos(10)},{sin(10)})$) {$\psi_s$};
% e_y
\draw[<-,color=blue] ($ (C) + 1.7*({cos(\psis)},{sin(\psis)}) $) -- ($ (C) + \lineRhoSBottom*({cos(-(90-\psis))},{sin(-(90-\psis))}) + 1.7*({cos(\psis)},{sin(\psis)}) $) node [pos=0.62, left=-0.1, color=blue] {$e_y$};
\node[color=red] (a) at (1.6,0.3) {$c_3$};
\node[color=red] (a) at (6.2,4.9) {$c_1$};
\node[color=red] (a) at (6.9,3.45) {$c_4$};
\node[color=red] (a) at (0.4,1.765) {$c_2$};
\end{tikzpicture}
\caption{A nonlinear dynamic bicycle model, including the representation of the curvilinear (road-aligned) coordinate system, and vehicle dimensions.}
\label{fig:spatial_bicycle_mdl}
\end{figure}
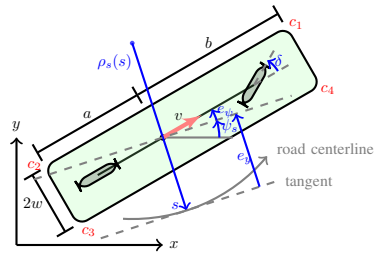

We address trajectory planning with obstacle avoidance. The main motivation and contribution is the development of a linear programming-based framework for the incorporation of actuator, obstacle avoidance, and, foremost, vehicle dimension constraints. A \emph{spatial-based} problem formulation is employed, eliminating time-dependency~\cite{kogan2006optimization,Gao:spatialmpc,frasch2013auto,verschueren2014towards,mogens:spatialcorridorplanning}. A sequential linear programming algorithm is motivated to successively improve planned trajectories.

Regarding obstacle avoidance, in~\cite{ziegler2014trajectory}, vehicle dimensions are accounted for by decomposing the vehicle shape into several circles of specific radius laid out equidistantly along the longitudinal vehicle axis. In~\cite{werling2012automatic}, safety distances are added to each side of the vehicle before making a hierarchical zero/one decision about interference with other obstacles. In~\cite{Werling:frenetframeoptimalplanning}, an obstacle proximity cost is considered. In~\cite{Gao:spatialmpc,frasch2013auto,mogens:spatialcorridorplanning} safety margins are added to obstacle contours such that point-mass trajectories can be planned; see also~\cite{lozano:spatialplanning} for a method transforming obstacle contours.

Regarding trajectory planning, in~\cite{fraichard2004reeds,vorobieva2014automatic,funke:lanechangeEP} reference paths composed of straights, arcs, and clothoids are used. In~\cite{Lima:MPCC}, a path sparsification method is presented that enables to fit a reduced number of clothoid segments to a reference path.   Alternative trajectory planning approaches include sampling-based methods, such as rapidly-exploring random trees (RRT)~\cite{lavalle:RRT}, B-splines~\cite{komoriya:bspline}, lattice-based motion planners~\cite{pivtoraiko:latticeplanners}, hybrid~\cite{Montemerlo:JuniorDARPA}, configuration-space planners~\cite{lozano:spatialplanning}, and hierarchical methods, such as~\cite{dolgov2010path}, where the output of a graph search based planner (A$^\star$) is consequently smoothed by a nonlinear optimization scheme to improve the quality of the solution. Though not addressing obstacle avoidance tasks, a sequential convex programming approach is employed in~\cite{dinh:timeoptSCP} when seeking the racing-line along a road segment. For a recent survey on motion planning, see~\cite{gonzalez2016review}; according to its taxonomy, the method presented in this paper can be classified as a \textit{numerical optimization approach}.
 
This paper is organized as follows. The problem formulation and main notation are defined in Section~\ref{sec:probform}. The SLP-Algorithm is stated in Section~\ref{sec_SCP}. Simulation results are reported in Section~\ref{sec_NumSim}, before concluding.

%%%%%%%%%%%%%%%%%%%%%%%%%%%%%%%%%%%%%%%%%%%%%%%%%%%%%%%%%%%%%
\section{Problem Formulation and Notation}
\label{sec:probform}

A path is sought avoiding obstacles, accounting for vehicle dimensions, traveling within road boundaries, respecting physical actuator constraints, and preferring smooth trajectories to increase safety by providing high maximum traveling speeds within friction limits. We consider two coordinate frames: a \textit{global} one within the $(x,y)$-plane and a \textit{road-aligned} one within the $(s,e_y)$-plane, see Fig.~\ref{fig:spatial_bicycle_mdl}. The road centerline coordinate is denoted by $s$. Let a trajectory planned at time $t$ be defined by $z(s) = \begin{bmatrix}  e_{\psi}(s) & e_{y}(s) \end{bmatrix}^T$, with $s\in[s_t,s_t+S]$, where the corridor length is $S>0$ and $s_t$ denotes the vehicle's position along the road centerline. The driving corridor is defined by spatially dependent convex bounds $e_{y}^\text{min}(s)\leq e_{y}(s) \leq e_y^\text{max}(s)$. \textit{Moving obstacles} are accounted for by a \textit{velocity-adjusted} mapping to the road-aligned coordinate system according to~\cite[Sect. III.E]{mogens:spatialcorridorplanning}. For few obstacles, a corridor may be determined based on heuristics (overtaking left or right). In general, a combinatorial problem has to be solved. In all of the following, this paper concentrates on the development of a trajectory planning method and assumes that a traversable corridor is \textit{given}. Throughout this paper \textit{forward motion} is assumed, i.e., $e_{\psi}(s)\in(-\frac{\pi}{2},\frac{\pi}{2})$. Let the associated trajectory in the $(x,y)$-plane be defined by $\mathcal{X}(s) = \begin{bmatrix} x(s) & y(s) & \psi(s) \end{bmatrix}^T$. The actual path length traveled by the vehicle is denoted by $\eta(s)$. It holds that ${\eta(s)=s}$ if ${e_\psi(s)=0}$ and ${e_y(s)=0,~\forall s}$. Otherwise, $\eta(s)\neq s$ because of lateral deviations of the vehicle's traveled path from the road centerline. For brevity, the distance argument is dropped in the following. We model vehicles as rectangles. This is a simple and yet an accurate vehicle representation. As illustrated in Fig.~\ref{fig:spatial_bicycle_mdl}, parameters $a$, $b$, and $w$ indicate distances between the center of gravity~(CoG) and rear, front and lateral vehicle sides, respectively. The four vehicle corners $c_i,~i=1,\dots,4$, can be expressed as
\begin{align}
s_{c_i} &= s + \xi_{c_i}^s \cos(e_\psi) + \zeta_{c_i}^s \sin(e_\psi),\label{eq_def_sci}\\
e_{y,c_i} &=  e_y +  \xi_{c_i}^{e_y} \cos(e_\psi) + \zeta_{c_i}^{e_y} \sin(e_\psi),\label{eq_def_eyci}
\end{align}
with $\xi_{c_i}^s,\xi_{c_i}^{e_y} \in\{b,-a,-a,b\}$, $\zeta_{c_i}^s\in\{-w,-w,w,w\}$, and $\zeta_{c_i}^{e_y}\in\{w,w,-w,-w\}$ for $c_i,~i=1,\dots,4$, respectively. Let the maximum and the rate steering actuator limitations be denoted by $\delta^\text{max}$ and $\dot{\delta}^\text{max}$. We assume symmetry, i.e., ${\delta^\text{min}=-\delta^\text{max}}$ and ${\dot{\delta}^\text{min}=-\dot{\delta}^\text{max}}$. Throughout this paper a goal is the minimization of absolute curvature peaks to maximize the lower bound on maximum permissible velocity within vehicle friction limits~\cite{funke:lanechangeEP}. Let the initial and desired end vehicle pose be indicated by $z(s_t)$ and $z(s_t+S)$.

For obstacle avoidance, this paper incorporates vehicle dimension constraints starting directly from nonlinear equations~\eqref{eq_def_sci} and~\eqref{eq_def_eyci}. Such constraints and the exploitation of available space is of particular relevance for maneuvering in tight spaces and for larger-sized vehicles.

%%%%%%%%%%%%%%%%%%%%%%%%%%%%%%%%%%%%%%%%%%%%%%%%%%%%%%%%%%%%%
\section{Sequential linear programming}\label{sec_SCP}

% ------------------------------------------------
\subsection{Spatial-based vehicle dynamics}\label{subsec_2dyn}

Consider the nonlinear kinematic bicycle model
\begin{equation}
\begin{bmatrix} \dot{x} & \dot{y} & \dot{\psi} \end{bmatrix}^T = \begin{bmatrix} v\cos(\psi) & v\sin(\psi) & \frac{v}{l}\tan(\delta) \end{bmatrix}^T,\label{eq_dotxypsi_nonlinkinbicmdl}
\end{equation}
assuming the CoG to be located at the rear axle and $l$ denoting the wheelbase. We abbreviate time and spatial derivatives by $\dot{x}=\frac{d x}{dt}$ and $x'=\frac{d x}{d s}$, respectively. In order to derive a spatial representation, we briefly review~\cite{mogens:spatialcorridorplanning}. In accordance with Fig.~\ref{fig:spatial_bicycle_mdl}, we have $\dot{e}_\psi=\dot{\psi}-\dot{\psi}_s$, ${\dot{e}_y=v \sin(e_\psi)}$, and ${\dot{s}=\frac{\rho_s v \cos(e_\psi)}{\rho_s-e_y}}$. Expressing ${e_\psi'=\frac{\dot{e}_\psi}{\dot{s}}}$ and ${e_y' = \frac{\dot{e}_y}{\dot{s}}}$, the spatial-based representation of~\eqref{eq_dotxypsi_nonlinkinbicmdl} is
\begin{equation}
\begin{bmatrix} e_\psi' & e_y' \end{bmatrix}^T \!=\! \begin{bmatrix} \frac{(\rho_s - e_y)\tan(\delta)}{\rho_s l \cos(e_\psi)} -\psi_s' & \frac{\rho_s-e_y}{\rho_s} \tan(e_\psi)  \end{bmatrix}^T.\label{eq_nonlinmdl_steering}
\end{equation}
The control variable is the front-axle steering angle $\delta$. Note that the spatial transformation eliminates any velocity-dependence in~\eqref{eq_nonlinmdl_steering}. This is characteristic for kinematic vehicle models but not the case for dynamic models~\cite{mogens:spatialcorridorplanning}. 

In~\cite{Lima:MPCC}, curvature $\kappa$ along the traveled vehicle path is related to steering angle $\delta$. For path sparsification, an $\ell_1$-optimization problem is then solved, where the decision variable is a set of discrete $\kappa$ expressed along a trajectory of waypoints. The optimized curvature sequence $\kappa$ is ultimately inverted and fed to a low-level feedback controller, which translates it to steering commands. It is well known that paths composed of clothoid concatenations are desirable and frequently employed in road design~\cite{Marzbani:RoadClothoids}. In a clothoid, the curvature varies linearly with the path arc-length. Thus, the curvature of paths composed of clothoid concatenations is PWA. Here an important remark can be made.
\begin{rmk}\label{Rmk1}
Expressing states and control variables as a function of path arc-length $s$ is advantageous when formulating linearly constrained optimization problems. This is since it is possible to formulate linear bounds on state variable $e_y$. These bounds are derived from, in general, spatially-varying road widths and coordinate transformation-distorted obstacles that can be approximated by their minimal rectangle-envelope within the $(s,e_y)$-plane. However, because of the resulting obstacle avoiding trajectory, the \textit{actual} planned path may significantly deviate from the road centerline.
 In accordance with Section~\ref{sec:probform}, the path arc-length along the \textit{actual} planned path is $\eta$. A PWA curvature profile $\kappa(\eta)$ would ensure clothoid-based path planning. However, a PWA $\kappa(s)$ does \textit{not} yield a clothoid path, unless $s=\eta$.
\end{rmk}

Let us denote~\eqref{eq_nonlinmdl_steering} by $z'=f(z,u)$ with $u=\delta$. Let a \textit{discretization grid} along the road centerline be defined by $\{s_j\}_{j=0}^N = \{s_0,s_1,\dots,s_N\}$, whereby for simplicity we abbreviated $s_j$ for $s_{t+j}$ when planning at time $t$. For a user-defined number of discretization points $N$, the discretization grid is initialized uniformly. New grid points are added such that all (potentially safety margin-adjusted) obstacle corners within the $(s,e_y)$-frame are accounted for. Consequently, the grid is, in general, non-uniformly spaced. Then, given a set of corresponding references $\{e_{\psi,j}^\text{ref}\}_{j=0}^N$, $\{e_{y,j}^\text{ref}\}_{j=0}^N$ and $\{u_j^\text{ref}\}_{j=0}^{N-1}$, the linearized and discretized system dynamics are ${z_{j+1} = A_jz_j + B_j u_j + g_j}$.

% ------------------------------------------------
\subsection{Linear vehicle dimension constraints}\label{subsec_vehDimCstrts}

Let us derive convex vehicle dimension constraints. At every $s_j$, assuming forward motion, we can describe lateral vehicle boundaries affine in $s$ and nonlinear in $e_{\psi,j}$ as
\begin{align}
e_{y,j}^\text{lower}(s) &= \tan(e_{\psi,j}) (s-s_{j,c_3}) + e_{y,c_3},\label{eq_eyjlower_nonlin}\\
e_{y,j}^\text{upper}(s) &= \tan(e_{\psi,j}) (s-s_{j,c_2}) + e_{y,c_2},\label{eq_eyjupper_nonlin}
\end{align}
\begin{figure}[t]
\vspace{0.25cm}
\centering
\input{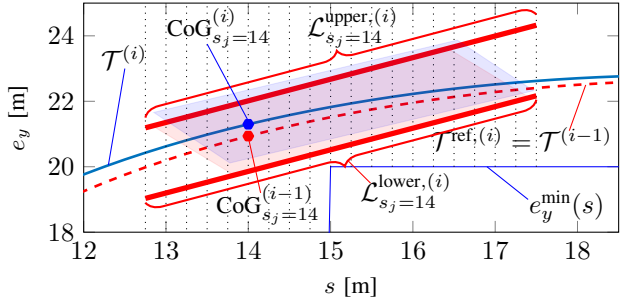}
\caption{Illustration of vehicle dimension constraints. Indices $(i)$ and $(i-1)$ indicate the corresponding SLP-iteration. The planning result at path coordinate $s_j=14$m is displayed.}
\label{fig_VehDimCstrt_plt1}
\end{figure}
accounting for~\eqref{eq_def_sci}. We define the set 
\begin{align} \tilde{\mathcal{S}}_j &= \{ \{s_k\}_{k=1}^{N-1}: s_j-\Delta s_{j,\text{min}} \leq s_k \leq s_j + \Delta s_{j,\text{max}} \}\nonumber\\ & =:\{s_{\tilde{k}_1},s_{\tilde{k}_2},\dots,s_{\tilde{k}_{\tilde{N}_j}}\},\nonumber
\end{align}
with 
$${\Delta s_{j,\text{min}}=\min(s_{j,c_2},s_{j,c_3})},{\Delta s_{j,\text{max}}=\max(s_{j,c_1},s_{j,c_4})},$$
and 
\begin{equation}
\mathcal{S}_j\!=\!\{\!s_{\tilde{k}_1-1},s_{\tilde{k}_1},\dots,s_{\tilde{k}_{\tilde{N}_j}},
s_{\tilde{k}_{\tilde{N}_j}+1}\}\! =:\! \{s_{k_1},\dots,s_{k_{\bar{N}_j}}\! \},\label{eq_def_mathcalSj}
\end{equation}
to also guarantee coverage of vehicle corners in between any two grid points. The linearization of~\eqref{eq_eyjlower_nonlin} and~\eqref{eq_eyjupper_nonlin} yields
\begin{align}
e_{y,\text{lin},j}^\text{lower}(s) &= \begin{bmatrix} g^\text{lower}(s) & 1 \end{bmatrix} \begin{bmatrix} e_{\psi,j} & e_{y,j} \end{bmatrix}^T + h_{\text{lin},j}^\text{lower}(s),\label{eq_eylinjlower_s}\\
e_{y,\text{lin},j}^\text{upper}(s) &= \begin{bmatrix} g^\text{upper}(s) & 1 \end{bmatrix} \begin{bmatrix} e_{\psi,j} & e_{y,j} \end{bmatrix}^T + h_{\text{lin},j}^\text{upper}(s),\label{eq_eylinjupper_s}
\end{align}
with $g^\text{lower}(s)$, $h_{\text{lin},j}^\text{lower}(s)$, $g^\text{upper}(s)$, and $h_{\text{lin},j}^\text{upper}(s)$ parameterized by $s_j$, $e_{\psi,j}^\text{ref}$, and $e_{y,j}^\text{ref}$. The main motivation of vehicle dimension constraints is to ensure that the vehicle geometry is constrained to the interior of the road corridor. By evaluating~\eqref{eq_eylinjlower_s} and~\eqref{eq_eylinjupper_s} at the discrete grid points of~\eqref{eq_def_mathcalSj}, this can be expressed as the set of inequalities
\begin{align}
\begin{bmatrix} e_{y,\text{lin},j}^\text{lower}(s_{k_1}) \\ \vdots \\ e_{y,\text{lin},j}^\text{lower}(s_{k_{\bar{N}_j}})  \end{bmatrix} &\geq \begin{bmatrix} e_y^\text{min}(s_{k_1}) \\ \vdots \\ e_y^\text{min}(s_{k_{\bar{N}_j}}) \end{bmatrix},\label{eq_eylinjlower_vec}\\
\begin{bmatrix} e_{y,\text{lin},j}^\text{upper}(s_{k_1}) \\ \vdots \\ e_{y,\text{lin},j}^\text{upper}(s_{k_{\bar{N}_j}})  \end{bmatrix} &\leq \begin{bmatrix} e_y^\text{max}(s_{k_1}) \\ \vdots \\ e_y^\text{max}(s_{k_{\bar{N}_j}}) \end{bmatrix}.\label{eq_eylinjlupper_vec}
\end{align}
We summarize the left-hand sides of the inequality signs by $\mathcal{L}_{s_j}^\text{lower}$ and $\mathcal{L}_{s_j}^\text{upper}$, respectively. For visualization, see Fig.~\ref{fig_VehDimCstrt_plt1}. Inequalities~\eqref{eq_eylinjlower_vec} and~\eqref{eq_eylinjlupper_vec} are linear in state $z_j$ at position $s_j$ and can be compactly summarized as $Q_j^\text{lower} z_j \geq q_j^\text{lower}$, and $Q_j^\text{upper} z_j \leq q_j^\text{upper}$, with $Q_j^\text{lower},\ Q_j^\text{upper}\in\mathbb{R}^{\bar{N}_j\times 2}$,  $q_j^\text{lower},\ q_j^\text{upper}\in\mathbb{R}^{\bar{N}_j}$, and $\bar{N}_j$ variable for each $s_j$ and dependent on references $e_{\psi,j}^\text{ref}$ and $e_{y,j}^\text{ref}$. Finally, note that instead of $\tilde{\mathcal{S}}_j$, as a \textit{least-conservative} variant, it could be differentiated between two grid segments $\tilde{S}_j^\text{lower}$ and $\tilde{S}_j^\text{upper}$ that are different for both lateral vehicle sides, instead of having one $\tilde{S}_j$ common to both.

% ------------------------------------------------
\subsection{Linear programming formulation}

We propose the following linear programming (LP):
\small
\begin{subequations}
\label{eq:SCP}
\begin{align}
%\min_{\{u_j\}_{j=0}^{N-1},\atop{\sigma,\sigma_{e_{\psi}}^N,\sigma_{e_{y}}^N}} &\ \  \max|u| + \lambda \max|D_1u| + W_\sigma(\sigma + \sigma_{e_{\psi}}^N + \sigma_{e_{y}}^N)\label{eq_OP_objFcn}\\
%
\min &\ \  \max|u| + \lambda \max|D_1u| + W_\sigma(\sigma + \sigma_{e_{\psi}}^N + \sigma_{e_{y}}^N)\label{eq_OP_objFcn}\\
\mathrm{s.t.} &\ \ z_0 = z(s_t),~u_{-1}=u(s_t-D_s), \\
&\ \ z_j = \begin{bmatrix} e_{\psi,j} & e_{y,j} \end{bmatrix}^T,~ j=0,\dots,N, \\
&\ \ z_{j+1} = A_j z_j + B_ju_j + g_j, \ j = 0,\dots,N-1, \label{eq:OP_zjp1_eq}\\
&\ \ e_{\psi}(s_t+S) - \sigma_{e_{\psi}}^N \leq e_{\psi,N} \leq  e_{\psi}(s_t+S)  + \sigma_{e_{\psi}}^N, \\
&\ \ e_y(s_t+S) - \sigma_{e_{y}}^N \leq e_{y,N} \leq  e_y(s_t+S) + \sigma_{e_{y}}^N, \\
&\ \ e_{y,j}^\text{min} - \sigma \leq e_{y,j} \leq  e_{y,j}^\text{max} + \sigma, \ j = 1,\dots,N, \label{eq:OP_ey_constrts}\\
&\ \ u^{\text{min}} \leq u_j \leq  u^\text{max}, \ j = 0,\dots,N-1, \\
&\ \ \Delta u^{\text{min}} \leq u_j-u_{j-1} \leq \Delta u^\text{max},  \ j = 0,\dots,N-1, \label{eq:OP_Deltau_cstrts}\\
&\ \ Q_j^\text{lower} z_j \geq q_j^\text{lower} - \sigma\mathbb{1}_{\bar{N}_j}, \ j=1,\dots,N,\label{eq:OP_Vjlower_cstrts}\\
&\ \ Q_j^\text{upper} z_j \leq q_j^\text{upper} + \sigma\mathbb{1}_{\bar{N}_j}, \ j=1,\dots,N,\label{eq:OP_Vjupper_cstrts}\\
& \ \ \sigma\geq 0,~\sigma_{e_{y}^N}\geq 0,~\sigma_{e_{\psi}^N}\geq 0,
\end{align}
\end{subequations} \normalsize
with decision variables $\{u_j\}_{j=0}^{N-1}$, $\sigma$, $\sigma_{e_{\psi}}^N$, and $\sigma_{e_{y}}^N$, and where $|\cdot|$  and $\mathbb{1}$ denote the absolute value and a column-vector of ones, respectively. LP~\eqref{eq:SCP} is solved repeatedly as discussed in the next Section~\ref{subsec_SCPalg}. The initial state is $z(s_t)$. The previous input $u_{-1}$ is relevant for rate constraints~\eqref{eq:OP_Deltau_cstrts}. The desired end pose is given by $z(s_t+S)$. Constant upper and lower bounds are $u^\text{min}$, $u^\text{max}$, ${\Delta u^\text{min}=\dot{u}^\text{min} T_s}$, and ${\Delta u^\text{max} = \dot{u}^\text{max}T_s}$, where time $T_s$ is related heuristically to the discretization grid taking reference speed into account. In general, $T_s$ may also account for a curved reference trajectory using a spatial transformation~\cite{mogens:spatialcorridorplanning}. Vehicle dimension constraints~\eqref{eq:OP_Vjlower_cstrts} and~\eqref{eq:OP_Vjupper_cstrts} were discussed in Section~\ref{subsec_vehDimCstrts}. All state constraints are softened by the introduction of slack variables $\sigma$, $\sigma_{e_\psi}^N$ and $\sigma_{e_y}^N$ to ensure feasibility of~\eqref{eq:SCP}. Matrix $D_1$ denotes the space-based first-order difference operator acting on vectorized input $u\in\mathbb{R}^{N\times 1}$. Let us motivate the objective function choice~\eqref{eq_OP_objFcn}. The first term is to minimize the maximum absolute curvature along traveled path; thereby simultaneously maximizing the lower bound on maximum admissible speed within vehicle friction limits. The second term $\max|D_1 u|$ is for input signal smoothing. To strongly penalize state constraint violations, we select a high scalar weight $W_\sigma=10^4$. A benefit of the proposed LP-based path planning is that additional constraints can easily be added. To enforce the overtaking of $L$ obstacles in \textit{parallel} (without specifying the lateral distance though), we may add
\begin{equation}
e_{\psi,l}^\text{obs} -\sigma_{e_\psi}^N \leq e_{\psi,j} \leq e_{\psi,l}^\text{obs} + \sigma_{e_\psi}^N,~\forall j\in\mathcal{J}_l^\text{obs},
\label{eq:epsij_parallelOA}
\end{equation}
for $l=1,\dots,L$, and where $${\mathcal{J}_l^\text{obs} = \{j: s_l^{\text{obs,b}} \leq s_j \leq s_l^{\text{obs,e}},~j=1,\dots,N\}},$$where $e_{\psi,l}^\text{obs}$ denotes the heading of the rectangle-envelope of obstacle $l$, located between $s_l^{\text{obs,b}}$ and $s_l^{\text{obs,e}}$. We reused slack variable $\sigma_{e_\psi}^N$ to not introduce a new decision variable. 

Let us elaborate on the LP~\eqref{eq:SCP}. All $u_j$ are box-constrained. All slack variables (including the ones when resolving the $\max$-terms in \eqref{eq_OP_objFcn}) are non-negative. Thus, all decision variables are constrained to a polyhedron with multiple  extreme points. This polyhedral set is not upper-bounded w.r.t. the slack variables. However, since the slack variables are only additively included in a $\text{min}$-objective, they will never cause unboundedness or infeasibility of \eqref{eq:SCP}. This further implies that \eqref{eq_OP_objFcn} attains a finite minimum. Then, by the \textit{Fundamental Theorem of Linear Programming} \cite{schrijver1998theory}, the minimum is attained at an extreme point of the polyedron. Since the \textit{simplex} method searches among these points and since they are finite numbered, \eqref{eq:SCP} is guaranteed to be solved within a finite number of iterations. For the design of a customized LP-solver, e.g., based on the \textit{simplex} method, a \textit{warm-start} is expected to be particularly useful for this search among extreme points of the polyhedron.

Closed-loop stability in the classical sense of linear systems and Lyapunov functions cannot be guaranteed. However, the spatial modeling, in particular, constraints \eqref{eq:OP_ey_constrts} (minus slack variables) in combination with the discussion in Section \ref{subsec_vmaxfric}  enforce vehicle operation within road boundaries and within the stable tire friction domain.

% ------------------------------------------------
\subsection{SLP-Algorithm}\label{subsec_SCPalg}

For obstacle constellations requiring larger steering maneuvers, transition dynamics~\eqref{eq:OP_zjp1_eq} and vehicle dimension constraints~\eqref{eq:OP_Vjlower_cstrts} and~\eqref{eq:OP_Vjupper_cstrts} are strongly dependent on underlying reference trajectories used for linearization and discretization. This is of particular relevance for the first initialization of references, for which we reconstruct $\{e_{\psi,j}^\text{ref}\}_{j=0}^N$ and $\{e_{y,j}^\text{ref}\}_{j=0}^N$ from a least-heading-varying PWA path avoiding all obstacles.
\begin{figure}[t]
\vspace{0.3cm}
\begin{center}
\begin{tikzpicture}[thick,scale=0.5, every node/.style={scale=0.5}]%§
\draw[->] (0, 0) -- (1,0);
\draw[->] (1.2, 0.2) -- ($ (1.2,0.2) + 1*({cos(45)},{sin(45)})$);
\draw[-,color=black!50,dotted] (0, 0) -- ($ (0,0.0) + 2.5*({cos(90)},{sin(90)})$);
\draw[-,color=black!50,dotted] (1.2, 0.2) -- ($ (1.2,0.2) + 2.5*({cos(135)},{sin(135)})$);
\draw[->,color=blue,solid] (-1.9, 2.4) -- ($ (0,0.0) + 2.5*({cos(90)},{sin(90)})$) -- ($ (1.2,0.2) + 2.5*({cos(135)},{sin(135)})$) -- (1.7, 2.1);
\draw[->] (6, 0.5) -- (7,0.5);
\draw[->] (7.2, 0.7) -- ($ (7.2,0.7) + 1*({cos(40)},{sin(40)})$);
\draw[-,color=black!50,dotted] (6, 0.5) -- ($ (6,0.5) + 0.8*({cos(90)},{sin(90)})$);
\draw[-,color=black!50,dotted] (7.2, 0.7) -- ($ (7.2,0.7) + 1.0*({cos(135)},{sin(135)})$);
\draw[->,color=blue,solid] (5.4, 1.3) -- ($ (6.0,0.5) + 0.8*({cos(90)},{sin(90)})$) -- ($ (7.2,0.7) + 1.0*({cos(135)},{sin(135)})$) -- (7.0, 1.8);
\end{tikzpicture}
\end{center}
\caption{Sketching the motivation for SLP-iterations. Large deviations between the trajectory output from the LP-solution and the reference trajectory can result in jaggedness (left). This undesired jaggedness can be alleviated by solving \eqref{eq:SCP} repeatedly causing optimization and reference trajectory to converge iteratively (right).}
\label{fig_reasonWhySLP}
\end{figure}
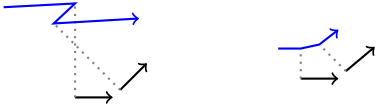
Therefore, we propose a sequential linear programming (SLP) approach, i.e., we sequentially solve~\eqref{eq:SCP} using as reference trajectory for linearization and discretization the solution of the previous SLP-iteration. See also Fig. \ref{fig_reasonWhySLP}. We initialize steering commands as $\{u_j^\text{ref}\}_{j=0}^{N-1}=0$. For our prototyping, the maximum admissible number of SLP-iterations is set as $I^\text{max}=5$. We distinguish between SLP and SLPp. The latter formulation additionally incorporates~\eqref{eq:epsij_parallelOA} into~\eqref{eq:SCP}. We summarize the following SLP-Algorithm:
\begin{enumerate}
\item Select: SLP or SLPp.
\item Initialize: $\{e_{\psi,j}^\text{ref}\}_{j=0}^N$, $\{e_{y,j}^\text{ref}\}_{j=0}^N$ and $\{u_j^\text{ref}\}_{j=0}^{N-1}$.
\item For $i\in\{1,\dots,I^\text{max}\}$:
\begin{itemize}
\item[-] Solve~\eqref{eq:SCP}; including~\eqref{eq:epsij_parallelOA} in case of SLPp.
\item[-] Update $\{e_{\psi,j}^\text{ref}\}_{j=0}^N$, $\{e_{y,j}^\text{ref}\}_{j=0}^N$ and $\{u_j^\text{ref}\}_{j=0}^{N-1}$.
\item[-] Check termination criterion.
\end{itemize}
\item Output: $\{e_{\psi,j}^\text{ref}\}_{j=0}^N$, $\{e_{y,j}^\text{ref}\}_{j=0}^N$ and $\{u_j^\text{ref}\}_{j=0}^{N-1}$.
\end{enumerate}
The termination criterion is as follows. Every $i$, we evaluate~\eqref{eq_def_sci} and~\eqref{eq_def_eyci} along $\{s_j\}_{j=0}^{N}$. If any obstacle or road boundary is hit or a jaggedness according to above sketch is encountered, an additional SLP-iteration is conducted.

\begin{figure*}[t!]
\vspace{0.3cm}
%\centering
\input{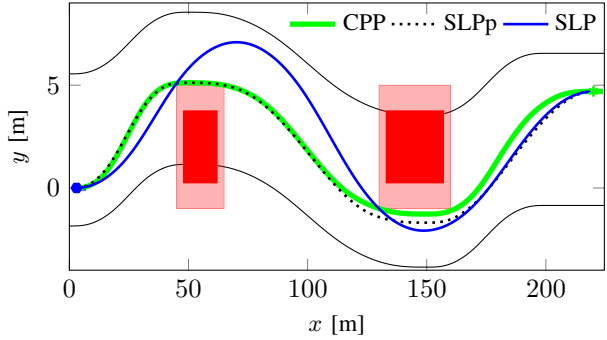}
\caption{Example 1. Comparison of the resulting trajectories in the global and in the road-aligned plane, in which computations are conducted. The obstacles (red) are inflated by a safety margin (light red). The light gray lines (right plot) indicate road bounds $e_y^\text{min}(s)$ and $e_y^\text{max}(s)$.}
\label{fig_TrajAllxysey_Ex1_Apr17}
\end{figure*}

% ------------------------------------------------
\subsection{Bounds on admissible traveling speeds}\label{subsec_vmaxfric}

The spatial transformation eliminates any velocity dependence of a kinematic vehicle model expressed in the road-aligned coordinate system, see Section~\ref{subsec_2dyn}. In order for our trajectory planning method to still provide velocity information, we employ the method from~\cite[Sect. 3]{funke:lanechangeEP} to determine \textit{spatially varying} upper bounds, $v^\text{max,fric}(\eta)$, on admissible traveling speeds. Besides the gravitational acceleration constant, a friction coefficient $\mu$ must be assumed. It is set as $\mu=0.8$ in subsequent simulations. Any reference traveling speeds $v^\text{ref}(\eta) \leq v^\text{max,fric}(\eta)$ are consequently within vehicle tire friction limits.

%%%%%%%%%%%%%%%%%%%%%%%%%%%%%%%%%%%%%%%%%%%%%%%%%%%%%%%%%%%%%
\section{Numerical results}\label{sec_NumSim}

For fast prototyping we employ MATLAB R2016b and CVX~\cite{cvx}. Its default settings and solvers are used, which come with a high numerical precision. It was found that a much reduced solver precision did not affect trajectory results, but sped up solver time. Two examples are discussed. For both, $N=200$ discretization points are employed, resulting in different \textit{adaptive} spatial discretization step sizes according to Section~\ref{subsec_2dyn}. It was found that the larger $N$ (the finer the discretization grid), the more tightly obstacles are avoided. The solver time can be greatly lowered by small $N$. While we retrieved acceptable trajectories for both examples for as low as $N=40$, we found that the computation of $v^\text{max,fric}(\eta)$ according to Section \ref{subsec_vmaxfric} was sensitive numerically and required ideally very smooth trajectories and thus large $N$. More illustrative plots for both examples can be found online in the extended version of this paper.

% ----------

%%%%%%%%%%%%%%%%%%%%%%%%%%%%%%%%%%%%%%%%%%%%%%%%%%%%%%%%%%%%%
\subsection{Concatenating clothoids for path planning}\label{sec_CPP}

For comparison, we consider a semi-analytical path planning method based on~\cite{funke:lanechangeEP}, where three path primitives (straights, arcs and clothoids) were concatenated to plan emergency lane changes up to the vehicle's friction limits. According to \cite{gonzalez2016review}, path planning methods most applied in real implementations by research groups worldwide  are \textit{interpolation}-based. Clothoids, together with B\'{e}zier and polynomial curves, belong to that class. Here, we treat the corners of safety margin-adjusted obstacles as waypoints. Safety margins are added to obstacle contours to account for vehicle dimensions. Planning based on path primitives requires to select multiple parameters, such as arc length, straight length and symmetric point fractions~\cite{funke:lanechangeEP}. These selections can significantly affect results. To maintain simplicity, we focus on clothoids, straights, and symmetric trajectory design when performing lane changes (no ``early'' or ``late'' steering -- see~\cite{funke:lanechangeEP} and the discussion of the symmetric point fraction). In the following, the comparative method is abbreviated as CPP (clothoid path planning). Clothoids can be fitted in either the $(x,y)$-, or $(s,e_y)$-domain before then requiring a retransformation to the $(x,y)$-domain. For the first example, for interest, we employ the latter method. For the second example, the $(x,y)$- and $(s,e_y)$-domains are identical (no curved road). Ultimately, by first-order discretization and inversion of~\eqref{eq_dotxypsi_nonlinkinbicmdl}, we reconstruct steering command $\delta(\eta)$ along the traveled path coordinate $\eta$.

\begin{figure}[t!]
\centering
\input{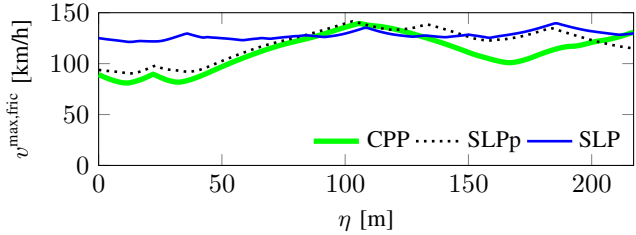}
\caption{Example 1. The maximum admissible traveling speed within vehicle tire friction limits is overall significantly higher for SLP in comparison to CPP and SLPp. For SLP, the lowest $v^{\text{max,fric}}(\eta)$ along its traveled path coordinate $\eta$ is $121$km/h. For CPP, the equivalent is 81km/h. Thus, the road segment could in principle be traversed much faster for SLP while still remaining in the safe vehicle tire friction domain.}
\label{fig_TrajAllvmaxfric_Ex1_Apr17}
\end{figure}

% ------------
% Ex2.

\begin{figure*}[t!]
\vspace{0.3cm}
\centering
\input{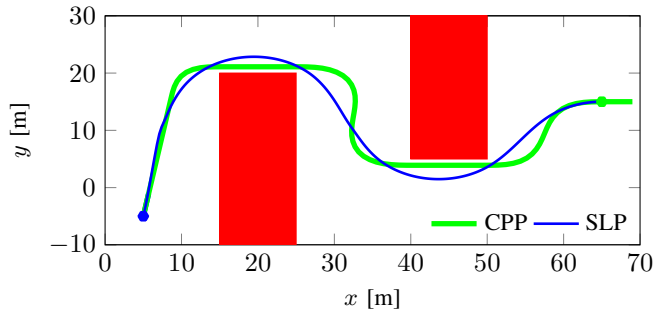}~
\input{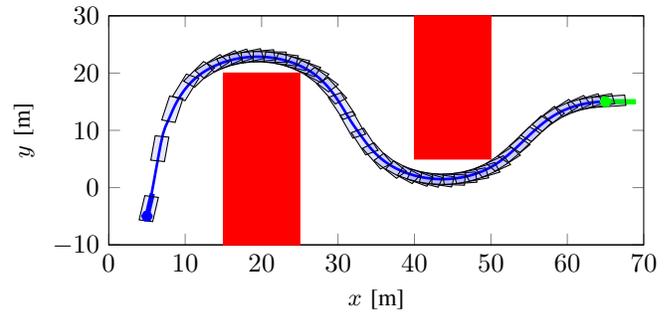}
\caption{Example 2. Resulting vehicle trajectories in the $(x,y)$-plane. To account for vehicle dimensions, trajectory planning with CPP assumed an obstacle inflated by a safety margin of 1.1m. For SLP, the right plot visualizes vehicle dimensions  (displayed every 5th sampling). According to the optimization problem formulation, obstacles are avoided tightly. In practice, small safety margins may additionally be added to the obstacle contours.}
\label{fig_Traj_all2_Ex2}
\end{figure*}

\begin{figure}[t!]
\centering
\input{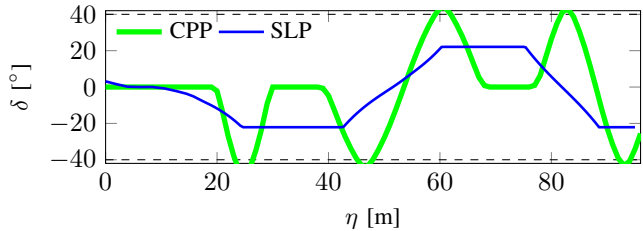}
\caption{Example 2. Actuator trajectories expressed along the traveled path coordinate $\eta$. The dashed lines indicate absolute actuator constraints. See Section~\ref{subsec_LowVelEx} for the discussion of constraint violations for CPP.}
\label{fig_TrajAlldelta_Ex2_Apr17}
\end{figure}

% ------------------------------------------------
\subsection{Example 1: roomy maneuvering space}

A curvy road profile with large inter-obstacle distance is considered (roomy maneuvering space). Specifically for a comparison between CPP and SLPp, we add safety margins to obstacles (admitting a safe overtaking in parallel) and dismiss constraints~\eqref{eq:OP_Vjlower_cstrts} and~\eqref{eq:OP_Vjupper_cstrts} from~\eqref{eq:SCP}. Results are visualized in Fig.~\ref{fig_TrajAllxysey_Ex1_Apr17}~and~\ref{fig_TrajAllvmaxfric_Ex1_Apr17}. Several observations can be made. First, as desired, SLP yields a trajectory that simultaneously maximizes the upper bound on maximal permissible velocity within friction limits. This comes, however, at the cost of reaching road boundary saturation and a \textit{non-parallel} overtaking of obstacles. Second, while SLPp overtakes the obstacles in parallel as desired, the solution of~\eqref{eq:SCP} with~\eqref{eq:epsij_parallelOA} produces a trajectory, that is laterally further displaced from the second obstacle in comparison to CPP (which is enforced to proceed along the waypoints). See Fig.~\ref{fig_TrajAllvmaxfric_Ex1_Apr17} for the effects on $v^\text{max,fric}$. Third, Fig.~\ref{fig_TrajAllxysey_Ex1_Apr17} depicts the distortions of obstacles, resulting from the spatial coordinate transformation from the $(x,y)$- to the $(s,e_y)$-plane. These distortions are minor for the given example. Note the difference between planned trajectories in the $(s,e_y)$- and $(x,y)$-plane after retransformation (from the curvilinear to the global coordinate system). While obstacles are overtaken in parallel in the $(s,e_y)$-plane for CPP and SLPp, the resulting $(x,y)$-trajectories avoid obstacles in a curved fashion after retransformation. Fourth, required steering actuation is confined to a region close to $0$. Fifth, for SLP two SLP-iterations were required. For SLPp, only one iteration was needed; this can be explained by the fact that the initial PWA reference trajectory according to Section~\ref{subsec_SCPalg} here already served as a sufficiently good reference.

% ------------------------------------------------
\subsection{Example 2: tight maneuvering space}\label{subsec_LowVelEx}

For a second example, a tight maneuvering space is assumed. Consider a planning scenario in a parking area. We incorporate constraints~\eqref{eq:OP_Vjlower_cstrts} and~\eqref{eq:OP_Vjupper_cstrts} into~\eqref{eq:SCP}. Results are visualized in Fig.~\ref{fig_Traj_all2_Ex2}~and~\ref{fig_TrajAlldelta_Ex2_Apr17}. Several observations can be made. First, steering commands of CPP, which were recomputed from planned trajectories according to Section~\ref{sec_CPP}, violate actuator constraints, see Fig.~\ref{fig_TrajAlldelta_Ex2_Apr17}. This has two causes: a) the absence of explicitly accounting for actuator constraints when trajectory planning according to CPP, and b) the enforcement of proceeding pairwise along waypoints, thereby also enforcing \textit{parallel} overtaking. In contrast, as Fig.~\ref{fig_TrajAlldelta_Ex2_Apr17} illustrates, SLP remains easily within actuation limits. This is enabled by the fact that SLP accounts for the entire obstacle constellation space (\textit{anticipative steering}). Second, a detail; as Fig.~\ref{fig_TrajAlldelta_Ex2_Apr17} illustrates, the final steering angle upon reaching the end pose is turned at -22.1$^\circ$. This is entirely in line with the formulation of~\eqref{eq:SCP} connecting the given start and end pose (and not planning beyond these). Third, four SLP-iterations were required overall to meet the termination criterion of Section~\ref{subsec_SCPalg}. Here a remark can be made. Note that it is distinguished between a) the $(s,e_y)$-frame and b) the reference trajectories $\{e_{\psi,j}^\text{ref}\}_{j=0}^N$, $\{e_{y,j}^\text{ref}\}_{j=0}^N$ and $\{u_j^\text{ref}\}_{j=0}^{N-1}$ on which linearized and discretized \eqref{eq:OP_zjp1_eq} as well as \eqref{eq:OP_Vjlower_cstrts} and \eqref{eq:OP_Vjupper_cstrts} are based on according to Section \ref{subsec_2dyn}. Furthermore, note that the linearization of \eqref{eq_nonlinmdl_steering} yields a pole at $e_{\psi,j}^\text{ref}= \pm 90^\circ,\forall j$. Accordingly, forward motion along positive $s$ is achieved only for $e_{\psi,j}\in(-\frac{\pi}{2},\frac{\pi}{2}),\forall j$. As Fig. \ref{fig_Traj_all2_Ex2} illustrates, the obstacle avoiding trajectory is initially exceeding $e_\psi>80^\circ$. Despite such a large deviation (close to the destabilizing $90^\circ$) from the road centerline, the SLP-algorithm is able to converge in only four iterations. This implies good robustness of the method with respect to the jaggedness-issue discussed in Section \ref{subsec_SCPalg}.

% ------------------------------------------------
\subsection{Discussion and limitations of the method}\label{subsec_limitations}

First, for \textit{road navigation} at most two SLP-iterations appear to be sufficient. This also holds for \textit{intersection navigation} since detailed maps (typical for autonomous vehicle applications) permit the formulation of suitable road centerlines that naturally avoid the problem of jaggedness discussed in Section \ref{subsec_SCPalg}. In contrast, for \textit{zone navigation} a different approach must be taken. Namely, a suitable ``road centerline'' must initially be set before two SLP-iterations can be applied. This initial setting of the road centerline, i.e., the definition of the $(s,e_y)$-frame,  also requires the mapping of all obstacles to it. In the simplest case, the initial road centerline can be set as a PWA as long as wedges exceeding the aforementioned $90^\circ$ are avoided.

Second, the formulation of~\eqref{eq:SCP} is based on a \textit{kinematic} vehicle model~\eqref{eq_dotxypsi_nonlinkinbicmdl}. \textit{Dynamic} vehicle models can be accounted similarly~\cite{mogens:spatialcorridorplanning}. It is not obvious to what extent this can improve safety. Rather than incorporating tire-dynamics in trajectory planning directly, the presented method simultaneously accounts for vehicle dimensions and plans trajectories according to a minmax objective involving steering angle; thereby already maximizing a stability measure in the form of maximized $v^\text{max,fric}(\eta)$. Any reference velocity with ${v^\text{ref}(\eta)\leq  v^\text{max,fric}(\eta)}$  ensures vehicle operation within friction limits (for a given $\mu$). Here, a reference velocity planning scheme may be based on~\cite{solea2006trajectory}. The employment of a kinematic model for trajectory planning bears the advantage of reduced computational complexity.

%%%%%%%%%%%%%%%%%%%%%%%%%%%%%%%%%%%%%%%%%%%%%%%%%%%%%%%
\section{Conclusion}\label{sec_concl}

We proposed a spatial-based trajectory planning method for automated vehicles. The main contribution was the incorporation of linearized vehicle dimension constraints within a sequential linear programming (SLP) algorithm and the relation of the proposed minmax-objective to increased bounds on admissible vehicle traveling speeds within tire friction limits. Anticipative steering accounting for the entire obstacle configuration space ranks among the main benefits. Future work comprises a) focus on zone navigation and the setting of a suitable initial $(s,e_y)$-frame, b) development of a customized LP-solver, and c) the application within a receding horizon control (RHC) scheme.

% ---------------------
%
%\balance % this command does the same thing as the next page! :-)
%
% -------------------------------
%\addtolength{\textheight}{-12cm}   % This command serves to balance the column lengths
                                  % on the last page of the document manually. It shortens
                                  % the textheight of the last page by a suitable amount.
                                  % This command does not take effect until the next page
                                  % so it should come on the page before the last. Make
                                  % sure that you do not shorten the textheight too much.

%%%%%%%%%%%%%%%%%%%%%%%%%%%%%%%%%%%%%%%%%%%%%%%%%%%%%%%%%%%%%%%%%%%%%%%%%%%%%%%%

%%%%%%%%%%%%%%%%%%%%%%%%%%%%%%%%%%%%%%%%%%%%%%%%%%%%%%%%%%%%%%%%%%%%%%%%%%%%%%%%

%%%%%%%%%%%%%%%%%%%%%%%%%%%%%%%%%%%%%%%%%%%%%%%%%%%%%%%%%%%%%%%%%%%%%%%%%%%%%%%%%
%\clearpage
%\section*{APPENDIX}
%Appendixes should appear before the acknowledgment.
%\section*{ACKNOWLEDGMENT}
%The preferred spelling of the word ÒacknowledgmentÓ in America is without an ÒeÓ after the ÒgÓ. Avoid the stilted expression, ÒOne of us (R. B. G.) thanks . . .Ó  Instead, try ÒR. B. G. thanksÓ. Put sponsor acknowledgments in the unnumbered footnote on the first page.
%%%%%%%%%%%%%%%%%%%%%%%%%%%%%%%%%%%%%%%%%%%%%%%%%%%%%%%%%%%%%%%%%%%%%%%%%%%%%%%%

% ----------------------------------------------------------
%\nocite{*}
%\bibliographystyle{plain}
\bibliographystyle{ieeetr}
\bibliography{bibref}

%%%%%%%%%%%%%%%%%%%%%%%%%%%%%%%%%%%%%%%%%%%%%%%%%%%%%%%%%%%
%% 
% UNCOMMENT FOLLOWING APPENDIX FOR ARXIV.
% 
% ----------------------------------------------

\section*{APPENDIX}
\subsection{Supplementary material}

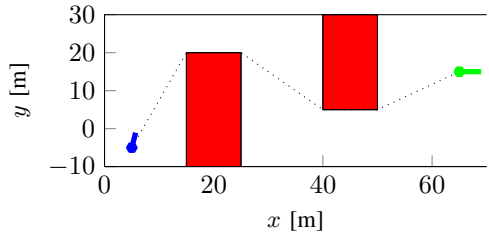
\begin{figure}[ht!]
\centering
% This file was created by matlab2tikz.
%
%The latest updates can be retrieved from
%  http://www.mathworks.com/matlabcentral/fileexchange/22022-matlab2tikz-matlab2tikz
%where you can also make suggestions and rate matlab2tikz.
%
\begin{tikzpicture}

\begin{axis}[%
width=2in,%2in,%4.521in,
height=0.8in,%1.5775in,%3.566in,
at={(0.758in,0in)},
scale only axis,
xmin=0,
xmax=70,
xlabel={\small{$x$ [m]}}, xlabel near ticks,
ymin=-10,
ymax=30.2,
ylabel={\small{$y$ [m]}}, ylabel near ticks,
axis background/.style={fill=white},
axis x line*=bottom,
axis y line*=left
]
\addplot [color=black,solid,forget plot]
  table[row sep=crcr]{%
1.83697019872103e-15	30\\
1	30\\
2	30\\
3	30\\
4	30\\
5	30\\
6	30\\
7	30\\
8	30\\
9	30\\
10	30\\
11	30\\
12	30\\
13	30\\
14	30\\
15	30\\
16	30\\
17	30\\
18	30\\
19	30\\
20	30\\
21	30\\
22	30\\
23	30\\
24	30\\
25	30\\
26	30\\
27	30\\
28	30\\
29	30\\
30	30\\
31	30\\
32	30\\
33	30\\
34	30\\
35	30\\
36	30\\
37	30\\
38	30\\
39	30\\
40	30\\
41	30\\
42	30\\
43	30\\
44	30\\
45	30\\
46	30\\
47	30\\
48	30\\
49	30\\
50	30\\
51	30\\
52	30\\
53	30\\
54	30\\
55	30\\
56	30\\
57	30\\
58	30\\
59	30\\
60	30\\
61	30\\
62	30\\
63	30\\
64	30\\
65	30\\
66	30\\
67	30\\
68	30\\
69	30\\
70	30\\
};
\addplot [color=black,solid,forget plot]
  table[row sep=crcr]{%
-6.12323399573677e-16	-10\\
0.999999999999999	-10\\
2	-10\\
3	-10\\
4	-10\\
5	-10\\
6	-10\\
7	-10\\
8	-10\\
9	-10\\
10	-10\\
11	-10\\
12	-10\\
13	-10\\
14	-10\\
15	-10\\
16	-10\\
17	-10\\
18	-10\\
19	-10\\
20	-10\\
21	-10\\
22	-10\\
23	-10\\
24	-10\\
25	-10\\
26	-10\\
27	-10\\
28	-10\\
29	-10\\
30	-10\\
31	-10\\
32	-10\\
33	-10\\
34	-10\\
35	-10\\
36	-10\\
37	-10\\
38	-10\\
39	-10\\
40	-10\\
41	-10\\
42	-10\\
43	-10\\
44	-10\\
45	-10\\
46	-10\\
47	-10\\
48	-10\\
49	-10\\
50	-10\\
51	-10\\
52	-10\\
53	-10\\
54	-10\\
55	-10\\
56	-10\\
57	-10\\
58	-10\\
59	-10\\
60	-10\\
61	-10\\
62	-10\\
63	-10\\
64	-10\\
65	-10\\
66	-10\\
67	-10\\
68	-10\\
69	-10\\
70	-10\\
};

\addplot[area legend,solid,draw=black,fill=red,forget plot]
table[row sep=crcr] {%
x	y\\
15	20\\
25	20\\
25	-10\\
15	-10\\
}--cycle;

\addplot[area legend,solid,draw=black,fill=red,fill opacity=0.3,forget plot]
table[row sep=crcr] {%
x	y\\
15	20\\
25	20\\
25	-10\\
15	-10\\
}--cycle;

\addplot[area legend,solid,draw=black,fill=red,forget plot]
table[row sep=crcr] {%
x	y\\
40	30\\
50	30\\
50	5\\
40	5\\
}--cycle;

\addplot[area legend,solid,draw=black,fill=red,fill opacity=0.3,forget plot]
table[row sep=crcr] {%
x	y\\
40	30\\
50	30\\
50	5\\
40	5\\
}--cycle;
\addplot [color=black,dotted,forget plot]
  table[row sep=crcr]{%
5	-5\\
15	20\\
25	20\\
40	5\\
50	5\\
65	15\\
};
\addplot [color=blue,line width=2.0pt,only marks,mark=asterisk,mark options={solid},forget plot]
  table[row sep=crcr]{%
5	-5\\
};
\addplot [color=blue,solid,line width=2.0pt,forget plot]
  table[row sep=crcr]{%
5	-5\\
5.69459271066772	-1.06076898795117\\
};
\addplot [color=green,line width=2.0pt,only marks,mark=asterisk,mark options={solid},forget plot]
  table[row sep=crcr]{%
65	15\\
};
\addplot [color=green,solid,line width=2.0pt,forget plot]
  table[row sep=crcr]{%
65	15\\
69	15\\
};
\end{axis}
\end{tikzpicture}%
\caption{Problem visualization.}
\label{fig_ProblVisualization}
\end{figure}

Problem formulation according to Fig. \ref{fig_ProblVisualization}: given a start pose (blue), a path is sought avoiding any obstacles (red), accounting for vehicle dimensions, traveling within corridor boundaries, respecting physical actuator constraints, and preferring smooth trajectories, thereby enabling high maximum traveling speeds within friction limits, such that an end pose (green) is reached. The black dotted path connects start and end pose traveling through obstacle corners in a piecewise-affine (PWA) fashion. It is infeasible to track by a real-world vehicle and, thus, requires smoothing: either spontaneously by direct tracking under actuator constraints, or, alternatively, using an explicit path planner preceding a tracking algorithm.

\begin{figure}[ht!]
\centering
\input{fig_TrajVehDimSLPP_Ex1_Apr17.tikz}
\caption{Example 1. The vehicle trajectories for SLP . Vehicle dimensions are visualized (displayed every 5th sampling).}
\label{fig_TrajVehDimSLPP_Ex1_Apr17_1}
\end{figure}

\begin{figure}[ht!]
\centering
\input{fig_TrajVehDimSLPPx_Ex1_Apr17.tikz}
\caption{Example 1. The vehicle trajectories for SLPp. Vehicle dimensions are visualized (displayed every 5th sampling).}
\label{fig_TrajVehDimSLPP_Ex1_Apr17_2}
\end{figure}

\begin{figure}[ht!]
\centering
\input{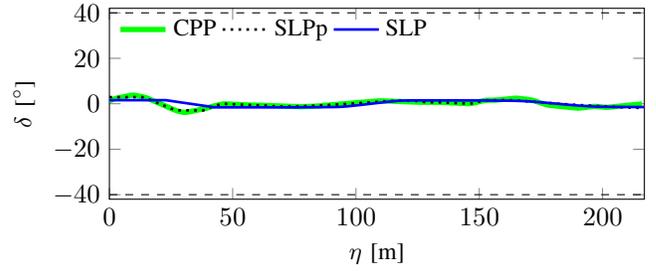}
\caption{Example 1. Resulting actuator trajectories. The dashed lines indicate absolute actuator constraints. For~$l=4.3$m, the bound $\delta^\text{max}=40^\circ$ corresponds to a minimal turning radius of $R_\text{min}=l/\tan(\delta^\text{max}) = 5.1$m. Because of the roomy maneuvering space, steering is confined closely to the origin, i.e., far from the actuation limits.}
\label{fig_TrajAlldelta_Ex1_Apr17}
\end{figure}

\begin{figure}[ht!]
\centering
% This file was created by matlab2tikz.
%
%The latest updates can be retrieved from
%  http://www.mathworks.com/matlabcentral/fileexchange/22022-matlab2tikz-matlab2tikz
%where you can also make suggestions and rate matlab2tikz.
%
\begin{tikzpicture}

\begin{axis}[%
width=2.8in,%1in,%1.952in,
height=1.0in,%3.566in,
at={(0in,0in)},
scale only axis,
xmin=0,
xmax=70,
xlabel={\small{$s$ [m]}}, xlabel near ticks,
ymin=-42,
ymax=42,
ylabel={\small{$\delta$ [$^\circ$]}}, ylabel near ticks,
axis background/.style={fill=white},
title style={font=\bfseries},
%title={Ctrls(s)},
%axis x line*=bottom,
%axis y line*=left,
legend style={at={(0,1)},anchor=north west,legend cell align=left,align=left,draw=white!15!black, draw=none, fill=none, legend columns=-1}
]
\addplot [color=black,dashed,forget plot]
  table[row sep=crcr]{%
0	40\\
70	40\\
};
\addplot [color=black,dashed,forget plot]
  table[row sep=crcr]{%
0	-40\\
70	-40\\
};
\addplot [color=blue,solid,line width=1.0pt]
  table[row sep=crcr]{%
5	3.11813083141557\\
5.3	1.65341064463443\\
5.6	0.188690058041988\\
5.9	0.00480235206434979\\
6.2	0.0427070857882371\\
6.5	-0.771308199207576\\
6.8	-2.23602998672048\\
7.1	-3.70075243794913\\
7.4	-5.16547509693269\\
7.7	-6.63019783079147\\
8	-8.09492058126449\\
8.3	-9.5596433186193\\
8.6	-11.0243660227169\\
8.9	-12.4890886721663\\
9.2	-13.9538112396536\\
9.5	-15.4185336851558\\
9.8	-16.8832559415498\\
10.1	-18.3479778770912\\
10.4	-19.8126991649032\\
10.7	-21.2774185009116\\
11	-22.0997485848701\\
11.3	-22.0997486809905\\
11.6	-22.0997487161299\\
11.9	-22.0997487443753\\
12.2	-22.0997487677839\\
12.5	-22.0997487876422\\
12.8	-22.0997488048096\\
13.1	-22.0997488198864\\
13.4	-22.0997488333054\\
13.7	-22.0997488453873\\
14	-22.0997488563748\\
14.3	-22.099748866455\\
14.6	-22.0997488757566\\
14.9	-22.0997417420672\\
15	-22.0997471022065\\
15.2	-22.0997488911481\\
15.5	-22.0997488988498\\
15.8	-22.0997489061251\\
16.1	-22.0997489130527\\
16.4	-22.0997489196804\\
16.7	-22.0997489260485\\
17	-22.0997489321916\\
17.3	-22.0997489381398\\
17.6	-22.0997489439199\\
17.9	-22.0997489495556\\
18.2	-22.0997489550685\\
18.5	-22.0997489604778\\
18.8	-22.0997489658018\\
19.1	-22.0997489710569\\
19.4	-22.0997489762589\\
19.7	-22.0997489814227\\
20	-22.0997489865626\\
20.3	-22.0997489916928\\
20.6	-22.0997489968272\\
20.9	-22.0997490019796\\
21.2	-22.0997490071641\\
21.5	-22.0997490123954\\
21.8	-22.0997490176886\\
22.1	-22.0997490230592\\
22.4	-22.0997490285244\\
22.7	-22.0997490341025\\
23	-22.0997490398135\\
23.3	-22.0997490456795\\
23.6	-22.0997490517249\\
23.9	-22.0997490579772\\
24.2	-22.0997490644667\\
24.5	-22.0997490712201\\
24.8	-22.0997474209391\\
25	-22.099742466309\\
25.1	-22.0997490842755\\
25.4	-22.0997490922532\\
25.7	-22.0997490933816\\
26	-22.099749060582\\
26.3	-22.0997490239485\\
26.6	-22.0997489823508\\
26.9	-21.9142798914662\\
27.2	-20.4495600532815\\
27.5	-18.9848384965469\\
27.8	-17.5201163702039\\
28.1	-16.0553939600966\\
28.4	-14.5906713804464\\
28.7	-13.1259486884255\\
29	-11.6612259168149\\
29.3	-10.1965030862411\\
29.6	-8.73178021061858\\
29.9	-7.26705729988272\\
30.2	-5.80233436151384\\
30.5	-4.33761140145018\\
30.8	-2.87288842467442\\
31.1	-1.40816543563465\\
31.4	0.0565575614183259\\
31.7	1.52128056213709\\
32	2.98600356183145\\
32.3	4.45072655532912\\
32.6	5.91544953702513\\
32.9	7.38017250106932\\
33.2	8.84489544137677\\
33.5	10.3096183511214\\
33.8	11.7743412216742\\
34.1	13.2390640410081\\
34.4	14.7037867911543\\
34.7	16.1685094433829\\
35	17.6332319474751\\
35.3	19.0979542031306\\
35.6	20.5626759598549\\
35.9	22.0273962162033\\
36.2	22.0997494802349\\
36.5	22.0997495390839\\
36.8	22.099749583812\\
37.1	22.0997496219807\\
37.4	22.0997496541071\\
37.7	22.0997496818307\\
38	22.099749706215\\
38.3	22.0997497279893\\
38.6	22.0997497476756\\
38.9	22.0997497656601\\
39.2	22.0997497822369\\
39.5	22.0997497976294\\
39.8	22.0997484945284\\
40	22.0997445714184\\
40.1	22.0997498245566\\
40.4	22.0997498374785\\
40.7	22.0997498496858\\
41	22.0997498613466\\
41.3	22.0997498725352\\
41.6	22.0997498833135\\
41.9	22.0997498937354\\
42.2	22.0997499038481\\
42.5	22.0997499136935\\
42.8	22.0997499233092\\
43.1	22.0997499327291\\
43.4	22.0997499419846\\
43.7	22.0997499511044\\
44	22.0997499601156\\
44.3	22.0997499690435\\
44.6	22.0997499779129\\
44.9	22.0997499867476\\
45.2	22.0997499955708\\
45.5	22.0997500044061\\
45.8	22.099750013277\\
46.1	22.0997500222078\\
46.4	22.0997500312234\\
46.7	22.0997500403497\\
47	22.0997500496146\\
47.3	22.0997500590481\\
47.6	22.0997500686827\\
47.9	22.0997500785542\\
48.2	22.0997500887023\\
48.5	22.0997500991715\\
48.8	22.099750110012\\
49.1	22.09975012128\\
49.4	22.0997501329769\\
49.7	21.8419325253168\\
50	20.3772114763469\\
50.3	18.9124892948447\\
50.6	17.4477667365326\\
50.9	15.9830439896638\\
51.2	14.5183211331134\\
51.5	13.0535982067323\\
51.8	11.5888752324929\\
52.1	10.1241522238739\\
52.4	8.65942918982046\\
52.7	7.194706136653\\
53	5.72998306908439\\
53.3	4.26525999080577\\
53.6	2.80053690484867\\
53.9	1.33581381382406\\
54.2	-0.128909279907791\\
54.5	-1.59363237410425\\
54.8	-3.05835546652415\\
55.1	-4.52307855481099\\
55.4	-5.98780163635906\\
55.7	-7.45252470814126\\
56	-8.91724776646503\\
56.3	-10.3819708066002\\
56.6	-11.8466938221701\\
56.9	-13.3114168040815\\
57.2	-14.7761397384848\\
57.5	-16.240862602449\\
57.8	-17.7055853533776\\
58.1	-19.1703078968187\\
58.4	-20.635029943148\\
58.7	-22.0997492652637\\
59	-22.0997507607984\\
59.3	-22.0997508167233\\
59.6	-22.09975086552\\
59.9	-22.0997509086818\\
60.2	-22.0997509472794\\
60.5	-22.0997509821268\\
60.8	-22.0997510138547\\
61.1	-22.09975104296\\
61.4	-22.0997510698401\\
61.7	-22.0997510948171\\
62	-22.0997511181551\\
62.3	-22.0997511400733\\
62.6	-22.0997511607555\\
62.9	-22.0997511803568\\
63.2	-22.09975119901\\
63.5	-22.0997512168293\\
63.8	-22.0997512339138\\
64.1	-22.0997512503504\\
64.4	-22.0997512662158\\
64.7	-22.0997512815781\\
};
\addlegendentry{SLP};

\end{axis}
\end{tikzpicture}%
\caption{Example 2. The result of the solution of~\eqref{eq:SCP}. Note that in contrast to Fig.~\ref{fig_TrajAlldelta_Ex2_Apr17}, the abscissa indicates the discretization coordinate $s$.}
\label{fig_TrajAlldeltaS_Ex2_Apr17}
\end{figure}
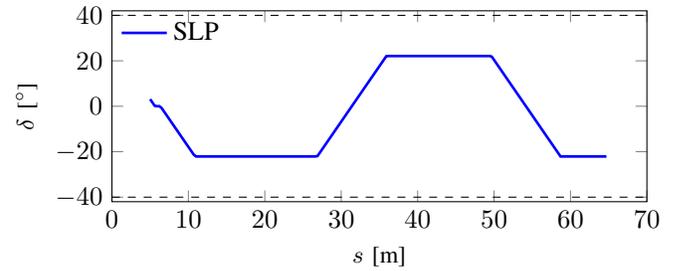

\begin{figure}[ht!]
\centering
% This file was created by matlab2tikz.
%
%The latest updates can be retrieved from
%  http://www.mathworks.com/matlabcentral/fileexchange/22022-matlab2tikz-matlab2tikz
%where you can also make suggestions and rate matlab2tikz.
%
\begin{tikzpicture}

\begin{axis}[%
width=2.8in,%1in,%1.952in,
height=1.0in,%3.566in,
at={(0in,0in)},
scale only axis,
xmin=0,
xmax=97,
xlabel={\small{$\eta$ [m]}}, xlabel near ticks,
ymin=0,
ymax=80,
ylabel={\small{$v^\text{max,fric}$ [km/h]}}, ylabel near ticks,
axis background/.style={fill=white},
title style={font=\bfseries},
legend style={at={(1,1)},anchor=north east,legend cell align=left,align=left,draw=white!15!black, draw=none, fill=none, legend columns=-1}
]
\addplot [color=green,solid,line width=2.0pt]
  table[row sep=crcr]{%
0	73.0041365510576\\
1	71.5973728118949\\
2	70.1624089777814\\
3	68.6974779272533\\
4	67.200619889741\\
5	65.6696516936512\\
6	64.1021293996187\\
7	62.495302491991\\
8	60.846057173538\\
9	59.1508454171659\\
10	57.4055951416363\\
11	55.605594984367\\
12	53.7453442966501\\
13	51.8183546011011\\
14	49.8168818129503\\
15	47.7315572086797\\
16	45.5508660023655\\
17	43.2603887357183\\
18	40.8416585555173\\
19	38.270365997276\\
20	35.5133883706618\\
21	32.5235390688876\\
22	29.2421694632345\\
23	25.9223508664197\\
24	23.0174532481953\\
25	20.7019028616325\\
26	19.4109871396799\\
27	19.4109871396799\\
28	22.1839067835498\\
29	24.9172562099188\\
30	27.9873214170822\\
31	31.2779364430345\\
32	34.3762922394854\\
33	37.2175983660218\\
34	39.8568662608403\\
35	42.3319022503656\\
36	44.6700135228825\\
37	45.5581599569693\\
38	43.2680688113625\\
39	40.8497933735876\\
40	38.2790472567997\\
41	35.5247017865293\\
42	32.6126106748841\\
43	29.7724730995531\\
44	27.1959090729478\\
45	24.9458116763519\\
46	23.0331198423701\\
47	21.4839355854025\\
48	21.3817921718507\\
49	21.3817921718507\\
50	23.6277146427547\\
51	25.6949986212282\\
52	27.993500340011\\
53	30.5602680015004\\
54	33.3476271535893\\
55	35.4449614106148\\
56	32.5094057319439\\
57	29.6768722343389\\
58	27.1117544802263\\
59	24.8743830865119\\
60	22.9755627598718\\
61	21.4298649272114\\
62	21.4298649272114\\
63	21.5688008118584\\
64	23.7243661782279\\
65	25.7829391800969\\
66	28.0875147756973\\
67	30.6623437198196\\
68	33.4524966138489\\
69	36.2952285969615\\
70	38.9949562257959\\
71	41.5214013618488\\
72	43.9026984484067\\
73	46.1613159588387\\
74	45.7649913313141\\
75	43.4857939050818\\
76	41.0803372862888\\
77	38.5249782810484\\
78	35.7876386433815\\
79	32.8287751860856\\
80	29.7115162130298\\
81	26.7718496667396\\
82	24.2606380052632\\
83	22.3502980819242\\
84	21.7505785479372\\
85	21.7505785479372\\
86	24.1202908249587\\
87	26.671868067372\\
88	29.5289293682095\\
89	32.5936416914315\\
90	30.9275529025843\\
91	27.8581345150228\\
92	25.1690668584732\\
93	23.0070246477137\\
94	21.6445228435315\\
95	21.6445228435315\\
96	23.4984478081888\\
97	25.7977607304128\\
};
\addlegendentry{\small{CPP}};

\addplot [color=blue,solid,line width=1.0pt]
  table[row sep=crcr]{%
0	67.6588258554917\\
1	66.1527916588622\\
2	64.6103680588483\\
3	63.406264405109\\
4	61.7818310227707\\
5	60.1786571747904\\
6	58.4640968403768\\
7	56.698232234479\\
8	54.8750447160962\\
9	52.9891550015495\\
10	51.0336206029246\\
11	49.0035111632355\\
12	46.8873013972513\\
13	44.6901536744002\\
14	42.4779223955509\\
15	40.5553627116464\\
16	39.7441381934678\\
17	39.7441381934678\\
18	42.2109922179756\\
19	43.5553104562751\\
20	41.5958914547764\\
21	39.826261382825\\
22	38.1412437323623\\
23	36.437517280839\\
24	35.0834686059097\\
25	34.0296902038954\\
26	33.3008551377515\\
27	33.3008551377515\\
28	32.7325485390633\\
29	32.7325485390633\\
30	32.8945386062439\\
31	32.8945386062439\\
32	32.7881621435515\\
33	32.7881621435515\\
34	33.0184583641593\\
35	32.5710223701619\\
36	32.5710223701619\\
37	32.9440586590912\\
38	32.8635782793629\\
39	32.7632261344203\\
40	32.7206965772539\\
41	32.7206965772539\\
42	32.8266642022551\\
43	32.8266642022551\\
44	33.8165267737354\\
45	35.1887642289866\\
46	36.8889576538986\\
47	38.8144609549434\\
48	40.870857673897\\
49	42.8841221936507\\
50	45.0631789776073\\
51	47.240105499819\\
52	49.3160027679829\\
53	47.8740028067473\\
54	46.2140179661533\\
55	43.991882070388\\
56	41.6270883528604\\
57	39.339591969183\\
58	37.2696252288558\\
59	35.5234041119924\\
60	34.1458348146319\\
61	33.2622118150922\\
62	33.2622118150922\\
63	32.9465087767956\\
64	32.9465087767956\\
65	32.9435102447558\\
66	32.9435102447558\\
67	33.0234805056017\\
68	33.0837867076131\\
69	32.5965279060976\\
70	32.5965279060976\\
71	33.0226364314846\\
72	32.9037594187839\\
73	32.7711189794704\\
74	32.7711189794704\\
75	32.7777105773194\\
76	32.7777105773194\\
77	34.2043640280997\\
78	35.9342907834131\\
79	37.9754778498746\\
80	40.1193163849708\\
81	42.3444086172335\\
82	44.6135576611024\\
83	44.8281577728845\\
84	42.5494636774418\\
85	40.296236868406\\
86	38.1301490300001\\
87	36.1014272499947\\
88	34.4878220097784\\
89	33.1701043116999\\
90	32.7886438669532\\
91	32.7886438669532\\
92	32.7886438669532\\
93	32.7886438669532\\
94	32.7886438669532\\
95	32.7886438669532\\
};
\addlegendentry{\small{SLP}};

\end{axis}
\end{tikzpicture}%
\caption{Example 2. The maximum admissible traveling speeds within vehicle tire friction limits. For SLP, the lowest $v^{\text{max,fric}}(\eta)$ along the traveled path coordinate $\eta$ is $32$km/h. For CPP, the equivalent is 19km/h. }
\label{fig_TrajAllvmaxfric_Ex2_Apr17}
\end{figure}
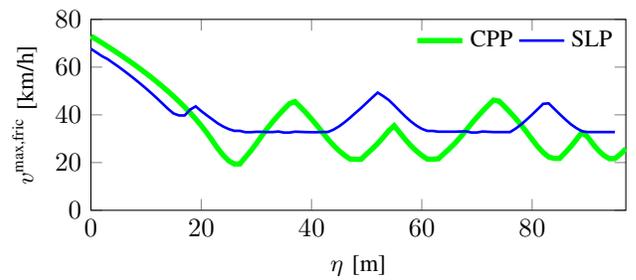

% ---------------------------------------------

\end{document}